\documentclass[format=acmsmall, review=false, screen=true]{acmart}

\setcopyright{acmlicensed}
\acmJournal{TEAC}
\acmYear{2017} \acmVolume{1} \acmNumber{1} \acmArticle{1} \acmMonth{1}
\acmDOI{http://dx.doi.org/10.1145/3033274.3084097}

\fancypagestyle{firstpagestyle}{
	\renewcommand{\headrulewidth}{0pt}%
	\fancyhf{}%
	\fancyfoot[L]%
}

\usepackage{booktabs} 
\usepackage{wrapfig}

\usepackage[ruled]{algorithm2e} 

\SetAlFnt{\small}
\SetAlCapFnt{\small}
\SetAlCapNameFnt{\small}
\SetAlCapHSkip{0pt}
\IncMargin{-\parindent}

\begin{document}
\title[Graphons for sparse Massive Networks]{Graphons:  A Nonparametric Method to Model, Estimate, and Design Algorithms for Massive Networks}
\titlenote{This extended abstract summarizes the content of an invited keynote talk delivered by Jennifer Chayes\ at the eighteenth ACM Conference on Economics and Computation (ACM EC'17).}
\author{Christian Borgs}
\affiliation{%
  \institution{Microsoft Research}
  \streetaddress{One Memorial Drive}
  \city{Cambridge}
  \state{MA}
  \postcode{02142}
  \country{USA}}
\author{Jennifer Chayes}
\affiliation{%
  \institution{Microsoft Research}
  \streetaddress{One Memorial Drive}
  \city{Cambridge}
  \state{MA}
  \postcode{02142}
  \country{USA}}

\ccsdesc[500]{Mathematics of computing~Graph theory}
\ccsdesc[500]{Mathematics of computing~Probabilistic inference problems}

\begin{abstract}
Many social and economic systems are naturally represented as networks, from off-line and on-line social networks, to bipartite networks, like Netflix and Amazon, between consumers and products.  Graphons, developed as limits of graphs, form a natural, nonparametric method to describe and estimate large networks like Facebook and LinkedIn.  Here we describe the development of the theory of graphons, for both dense and sparse networks, over the last decade.  We also review theorems showing that we can consistently estimate graphons from massive networks in a wide variety of models.  Finally, we show how to use graphons to estimate missing links in a sparse network, which has applications from estimating social and information networks in development economics, to rigorously and efficiently doing collaborative filtering with applications to movie recommendations in Netflix and product suggestions in Amazon.
\end{abstract}

\keywords{Networks; graphons; graph convergence; nonparametric estimation; collaborative filtering; network completion}

\maketitle

\fancypagestyle{acmec}{
	\renewcommand{\headrulewidth}{0pt}
	\fancyhf{}
	\fancyhead[C]{\footnotesize{\mbox{\shortauthors}}}
}
\pagestyle{acmec}
\renewcommand{\footnotetextcopyrightpermission}[1]{\thankses}


Networks are the natural framework to describe social and economic systems with pairwise interactions. This includes not only social networks among individuals, like Facebook and LinkedIn, but also bipartite networks between different types of entities, like the Netflix movie-user network and the Amazon consumer-product network. Graphons, developed as limits of sequences of graphs, form a natural, non-parametric method to model and estimate large networks. Here we informally describe some of the development of the theory of graphons over the last dozen years, and provide an application of graphons to the completion of noisily, sparsely sampled massive networks.

\section{Graph Limits}
The theory of limits of sequences of dense graphs (i.e., graphs in which the number of edges scale as the square of the number of vertices) was developed over roughly a decade starting around 2006 \cite{BCLSV06,LS,BCLSV2,BCLSV3}. Of course, it is always possible to a define a limit of a graph sequence -- if we make the notion too weak, then every sequence trivially converges to the same point; if we make it too strong, each sequence could converge to a different point. The key was to find a notion that was ``just right.'' This was done in a series of papers with Lov{\'a}sz, Sos and Vestergombi \cite{BCLSV2,BCLSV3}.

In \cite{BCLSV06} we define a metric to allow us to compare graphs of different sizes. For this, we rescale two graphs to be of the same size by creating copies of each vertex so that the two graphs then have the same number of vertices. Specifically, if the two graphs have $n_1$ and $n_2$ vertices, and $k_1$ and $k_2$ are the smallest numbers such that $k_1 n_1=k_2 n_2$, we ``blow up'' each vertex in the first graph by $k_1$, and each one in the second by $k_2$.  We then replace each edge in the original graphs by complete bipartite graphs between the blown-up vertices.  We look at the Frieze-Kannan cut distance \cite{FK99} of the difference between the two blown-up graphs, where the cut distance is basically the value of the cut between a set and its complement, taking the sup over all sets. Finally, since vertex labels should not matter, we take the minimum over all relabelings of the vertices. The resulting metric is called the cut metric.

Our first notion of convergence \cite{BCLSV06,BCLSV2} is what we call metric convergence, which means simply that the sequence of graphs is Cauchy in the cut metric. The fact that a limit exists was established by Lov{\'a}sz and Szegedy \cite{LS} using a weak version of the Szemer\'edi Regularity Lemma \cite{FK99} and a martingale argument. The limit is a two-variable symmetric function, called a graphon, often denoted by $W(x,y)$, which is basically a continuous version of the adjacency matrix of a graph on ${[0,1]}^2$. In \cite{BCL}, we show that the limit is unique up to measure-preserving transformations of the underlying feature space.

\begin{figure}
\begin{center}
\includegraphics[scale=.4]{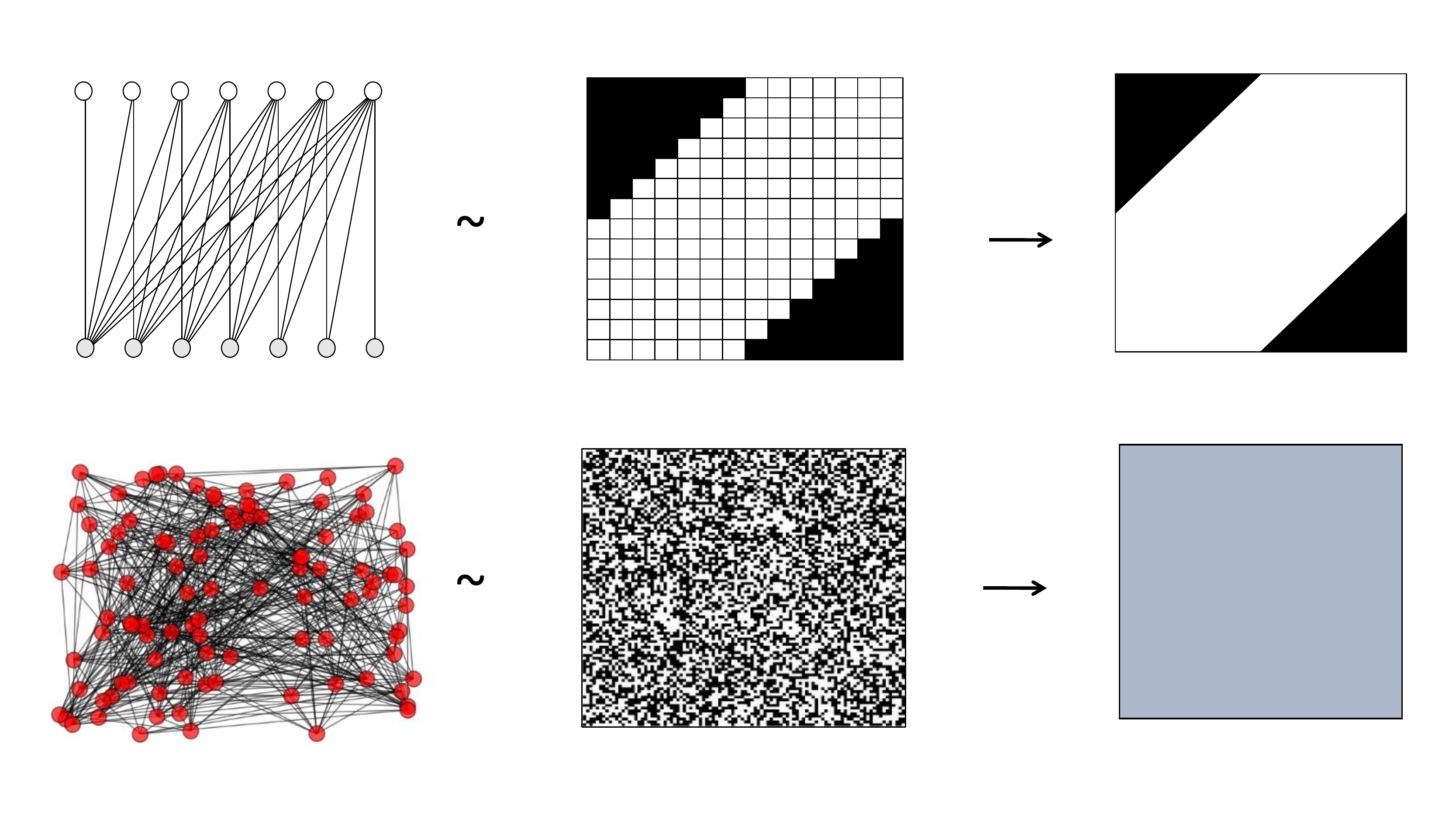}
\end{center}
\caption{The half-graph and a random graph, their empirical graphons, and their limits.}
\label{fig1}
\end{figure}

Note that a graph on $n$ vertices can be represented as a piecewise constant graphon by dividing the unit square ${[0,1]}^2$ into $n^2$ small squares of side length $1/n$, and representing the presence of an edge between vertices $i$ and $j$ by a $1$ on the squares $ij$ and $ji$, and its absence by a $0$.  This representation is called the empirical graphon of the graph. If we visualize a $0$ in the empirical graphon as a small white square, and a $1$ as a small black square, graph convergence can be seen as convergence of the these black and white empirical graphons to a gray-scale limiting graphon.  See Figure~\ref{fig1} for an illustration for the half-graph and an Erd\"os-Reny\'i
random graph.

There is a natural local notion of convergence in dense graphs.  We can ask that all induced finite subgraph densities (e.g., the edge density, triangle density, Peterson graph density, \dots) converge simultaneously. We prove that a sequence of dense graphs converges in metric if and only if it is subgraph convergent \cite{BCLSV2}.

We also introduce several ``global'' notions of convergence \cite{BCLSV3, BCG}. We have one notion which corresponds to convergence of all finite two-body statistical physics models in the microcanonical ensemble (the ensemble in which each species of vertex represents a fixed percentage of all vertices). We have another notion which corresponds to all multi-way cuts and bisections converging.  We show that a sequence of dense graphs converges according to these and a few other global notions if and only if it converges in metric \cite{BCLSV3}. More recently, a notion of large deviations convergence was defined \cite{BCG}, which is also equivalent with the other global and local notions for dense graphs.

Sparse graphs of bounded degree are much more fragile. Here we have no notion of metric convergence. The natural notion of local convergence is convergence of a graph rooted at a random point, known as Benjamini-Schramm convergence \cite{BS}.  Benjamini-Schramm convergence is equivalent to subgraph convergence \cite{BCKL}, but global notions of convergence are stronger \cite{BCG}.

The most interesting case is that of sparse graphs of unbounded average degree, like the power-law networks in Facebook and LinkedIn. These cases were much more complicated due to the fact that much of the necessary graph theory (e.g., the relevant case of the weak Regularity Lemma) had not yet been developed.  Over the past five years, we have developed two ways to generalize the theory.

The first way is by rescaling the empirical graphon function, dividing it by the edge density $\rho$ of the graph sequence (which is tending to zero in the case of sparse graphs). This was done by Bollob\'as and Riordan in the case when the graphs had no ``dense spots,'' corresponding to the graphon being a bounded function \cite{BR}. In collaboration with Cohn and Zhao \cite{BCCZ1,BCCZ2}, we develop a theory of ``uniformly upper regular'' sparse graphs of unbounded average degree; we show that these converge to graphons in $L^p$ spaces, corresponding to growing sequences with power-law tails. In the process of doing this, we formulate and prove a weak form of the Regularity Lemma for sparse graphs corresponding to graphons of unbounded degree.

\begin{figure}
\begin{center}
\includegraphics[scale=.5]{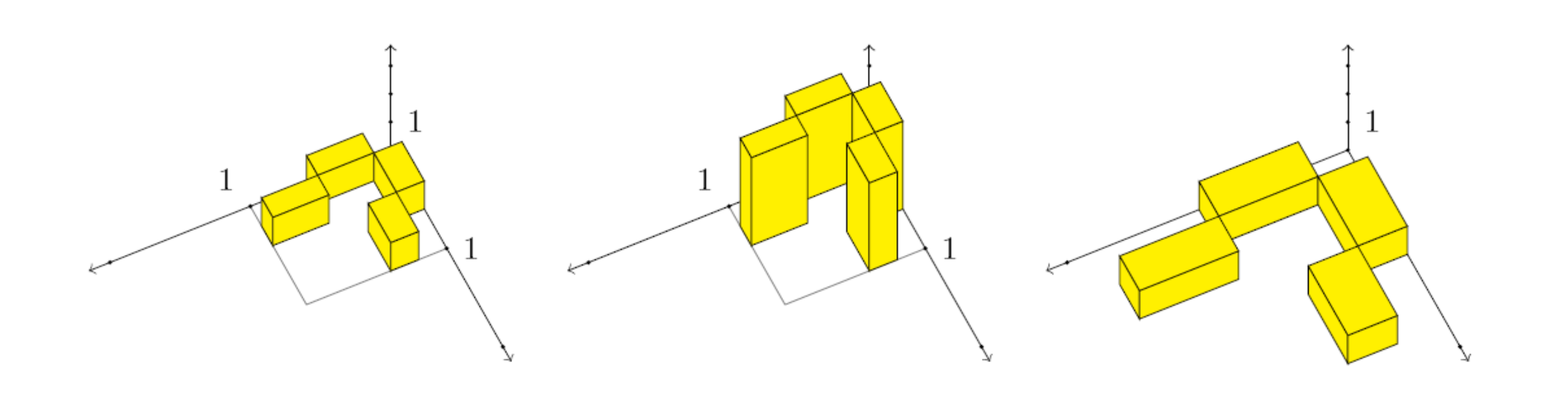}
\end{center}
\caption{Three graphons associated with the same simple
graph $G$ on five vertices: the emprical graphon used in theory of dense
graph convergence, the rescaled a version used in
the uniformly upper regular theory of sparse graph convergence,
and the stretched version used in the theory for sparse graphs with uniformly regular tails.}
\label{fig2}
\end{figure}

The second way to get graph limits for graphs of unbounded average degree, again graphs with very long tails, is by ``stretching'' the empirical graphon (and its domain) by $1/ \sqrt{\rho}$ so that the integral of the graphon is again unity. This  leads to a limit defined on the positive quadrant rather than the unit square, and hence these graphons provide a good description of networks where the underlying features cannot be restricted to bounded domains.  See Figure~\ref{fig2} for a comparison of this with the rescaling approach. In work with Cohn and Holden \cite{BCCH}, we develop a theory of sparse graphs of unbounded degree with ``uniformly regular tails,'' show how to construct limits for sequences obeying this regularity condition, and prove many properties of the these graphons.  Finally, we show that any graph sequence which satisfies both the conditions of uniform upper regularity and uniformly regular tails must be dense; so dense graphs are the only overlap of our two theories of sparse graphs of unbounded degree.

\section{Modeling Large Networks}

The classic way to model large networks  uses the so-called stochastic block model
\cite{HLL}, which is a generalization of the Erd\"os-Reny\'i random graph to $k$ species of vertices, with fixed densities of each species of vertex, and fixed probabilities that any two species connect to each other, including a fixed probability of intra-species connection, which is in general different within each species.

As graphs get larger and larger, reasonable descriptions have the number of species scale with the number of vertices: $k=k(n)$, which tends to lead to overfitting.  Instead, starting over a decade ago \cite{HRH}, researchers began to model large networks non-parametrically by choosing $n$ points randomly from some feature space, $x_1, x_2, \cdots,  x_n \in \Omega$, and a two-variable, bounded symmetric function $W(x_i,x_j): \Omega \times \Omega \rightarrow [0,1]$. A random graph consistent with this function is then generated by saying that $i$ is connected to $j$ with probability $W(x_i,x_j)$.

The graphs generated in this way can easily be seen to be dense, whereas real
life networks tend to be sparse.  But  our  two methods for constructing graph limits for sequences of random sparse graphs of unbounded degree naturally yield random models of massive sparse networks.

Our first method \cite{BCCZ1} chooses features $x_1, x_2, \cdots,  x_n$ from some compact feature space, and generates random graphs with long tails corresponding to unbounded graphons $W$, connecting vertices with features $x_i$ and $x_j$ with probability
$\min\{1,\rho W(x_i,x_j)\}$, where $\rho$ is the desired target density of our graph.

Our second method \cite{BCCH} is more complex:  It chooses features $x_1, x_2, \cdots $ from an unbounded feature space according to a Poisson process with an intensity which scales linearly with time. At time $t$, the process connects   vertices $i$ and $j$ born up to time $t$ with probability $W(x_i,x_j)$. The resulting graph at time $t$ retains only those vertices which have been connected to at least one other vertex by that time, leading to a graph with a finite number of vertices at any finite time.
We show that this kind of model is quite natural by proving a generalization of the celebrated Aldous-Hoover Theorem \cite{A,H} to sparse graphs. In our setting,
it says that under suitable regularity conditions, any family of graphs with vertices labeled by their birth time that is invariant under measure-preserving transformations of time must be of the above form.  Our  model  generalizes the work of Caron and Fox \cite{CF}, and is very similar to a model introduced
simultaneously by Veitch and Roy \cite{VR}.

\section{Learning Sparse Graphs}

How does one learn a network from a single sample, e.g., allowing us to predict how the network might look when it's twice its current size? The traditional approach is to assume that the network is a stochastic block model with $k$ types of vertices, cluster the vertices according to type, and then use this clustering to estimate the parameters of the model (the densities of each of the $k$ types, plus the $k(k+1)/2$ connection probabilities within and between types)
and then use that estimated stochastic block model to make predictions about the network
\cite{WBB76,HLL,FMW85,WW87,SN97,JS98,B87,McSherry01,DHKM06,C-O10,CCT12,AGHK14}.
See \cite{AS15,AS15b} for recent work with rigorous consistency guarantees for
 sparse graphs.

 As we discussed above, for very large networks, it is often better to take a non-parametric approach \cite{Kal99,BC09,RCY11,CWA12,BCL11}. Here the idea is to get an estimate, $\widehat W$ for the graphon $W$ representing the network, and then generate realizations from that estimation. For bounded graphons, it had been shown that this leads to consistent estimation for both dense and sparse graphs,
  see, e.g., \cite{C,GLZ15,WO,KTV15}, to name a few of the more recent papers on the subject.
  In collaboration with Cohn and Ganguly \cite{BCCG}, we showed that this leads to consistent estimation for sparse massive networks if, e.g., $W \in L^2$, provided that the average degree diverges.  Hence one can use graphons for statistically consistent estimation for power-law graphs. In collaboration with Smith \cite{BCS}, we showed that, for sparse graphs corresponding to bounded $W$, it is possible to consistently estimate while maintaining edge differential privacy.

\section{Network Completion and Collaborative Filtering on Sparse Graphs}
An area of some interest in economics is how to ``complete'' sparsely, noisily sampled networks. For example, in development economics, one sometimes has measurements representing sparsely and noisily sampled connections among individuals. How sparsely can we sample and still get an accurate estimate of the missing links? Similarly, in bipartite networks like the Netflix user-movie network or the Amazon consumer-product network, can we show that we can complete the network even if it is very sparsely sampled with some general noise?  Here we follow a collaborative filtering approach, which however must be modified to account for the sparsity of the sample.
What if the sampling is so sparse that two people for whom we are trying to estimate the probability of a connection have no connections in common? Or what if there are no sets of movies in common between a given individual and another on the Netflix network? In collaboration with Lee and Shah \cite{BCLS}, we used the ``expanded neighborhood'' approach of Abbe and Sandon \cite{AS15,AS15b}, who study estimation for stochastic block models. The idea is to look not just at individuals who share a common connection, but instead expand the neighborhood until a common connection is found. Similarly, we expand the bipartite user-movie-user-movie-$\cdots$ neighborhood until there  an overlap. Here we treat graphons as operators in the expansion and need to do estimates to control the variance. Assume there are $n$ individuals on the network, each described by features in a $d$-dimensional space. We show that, if the underlying graphon describing this network is Lipshitz, then the mean square error of our estimated graphon tends to zero provided that we sample the network with density $p=\omega(d^2 n^{-1})$. For high-dimensional latent spaces, which tend to be the appropriate descriptions for most real-world networks, this is a significant improvement over the best previous result, that of Chatterjee \cite{C}, who showed that the mean square error of the estimated graphon tends to zero provided that the network is sampled with density $p=\omega( n^{-2/(d+2)})$.

\bibliographystyle{ACM-Reference-Format}
\bibliography{BC-EC17}

\end{document}